\documentclass[a4paper,fleqn]{cas-dc}

\usepackage[numbers]{natbib}
\usepackage{lineno}
\usepackage{graphicx}
\usepackage{amsfonts}
\usepackage{gensymb}
\usepackage{xspace} 

\def\tsc#1{\csdef{#1}{\textsc{\lowercase{#1}}\xspace}}
\tsc{WGM}
\tsc{QE}
\tsc{EP}
\tsc{PMS}
\tsc{BEC}
\tsc{DE}

\begin{document}
\let\WriteBookmarks\relax
\def\floatpagepagefraction{1}
\def\textpagefraction{.001}

\shorttitle{The absolute energy scale of the DAMPE}

\shortauthors{J. Zang et~al.}

\title [mode = title]{Determination of the absolute energy scale of the DAMPE calorimeter with the
geomagnetic rigidity cutoff method}



%

\author[add1,add2]{JingJing~Zang}[orcid=0000-0002-2634-2960]
\ead{zangjingjing@lyu.edu.cn}

\author[2]{Chuan~Yue}[orcid=0000-0002-1345-092X]
\cormark[2]
\ead{yuechuan@pmo.ac.cn}
\cortext[2]{Corresponding author}

\author[add2,add6]{Qiang~Yuan}
\author[add2]{Wei~Jiang}
\author[add2]{Xiang~Li}
\author[add3,add4]{Yunlong~Zhang}
\author[add3,add4]{Cong~Zhao}
\author[add5]{Fabio~Gargano}
\affiliation[add1]{organization={School of Physics and Electronic Engineering, Linyi University, Linyi 276000, China}}
\affiliation[add2]{organization={Key Laboratory of Dark Matter and Space Astronomy, Purple Mountain Observatory, Chinese Academy of Sciences, Nanjing 210023, China}}
\affiliation[add6]{organization={School of Astronomy and Space Science, University of Science and Technology of China, Hefei 230026, People’s Republic of China}}
\affiliation[add3]{organization={State Key Laboratory of Particle Detection and Electronics, University of Science and Technology of China, Hefei 230026, China}}
\affiliation[add4]{organization={Department of Modern Physics, University of Science and Technology of China, Hefei 230026, China}}
\affiliation[add5]{organization={Istituto Nazionale di Fisica Nucleare, Sezione di Bari, via Orabona 4, I-70126 Bari, Italy}}


\begin{abstract}
The Dark Matter Particle Explorer (DAMPE) is a satellite-borne detector designed to detect high-energy cosmic ray particles with its core component being a BGO calorimeter capable of measuring energies from $\sim$GeV to $O(100)$ TeV.
The 32 radiation lengths thickness of the calorimeter is designed to ensure full containment of showers produced by cosmic ray electrons and positrons (CREs) and $\gamma$-rays at energies below tens of TeV, providing high resolution in energy measurements. The absolute energy scale therefore becomes a crucial parameter for precise measurements of the CRE energy spectrum. The geomagnetic field induces a rapid drop in the low energy spectrum of electrons and positrons, a phenomenon that provides a method to determine the calorimeter's absolute energy scale. By comparing the cutoff energies of the measured spectra of CREs with those expected from the International Geomagnetic Reference Field model across 4 McIlwain $L$ bins - which cover most regions of the DAMPE orbit - we find that the calorimeter's absolute energy scale exceeds  the calibration based on Geant4 simulation by $1.013\pm0.012_{\rm stat}\pm0.026_{\rm sys}$
for energies between 7 GeV and 16 GeV. The absolute energy scale should be taken into account when comparing the absolute CREs fluxes among different detectors.
\end{abstract}




\begin{keywords}
absolute energy scale \sep geomagnetic rigidity cutoffs \sep cosmic rays electrons \sep and positrons \sep \WGM \sep \BEC
\end{keywords}

\maketitle


\section{Introduction} \label{sec:intro}

The Dark Matter Particle Explorer (DAMPE) is a satellite experiment designed for the detection
of high-energy cosmic rays and $\gamma$-rays, to explore the acceleration and propagation
physics of cosmic rays and search for possible new physics signals from dark matter
annihilation or decay. It was launched at the end of 2015, and has operated continuously in space for over 9 years. The payload consists of four sub-detectors: the plastic scintillator
detector (PSD), the silicon tungsten tracker (STK), the BGO calorimeter (BGO), and the neutron
detector (NUD). Precise measurements of the charge, directions and energy of high-energy
particles are achieved through the combined operation of the four sub-detectors, advancing our understanding of the high-energy Universe \citep{Chang2017b,
2019SciA....5.3793A, 2021PhRvL.126t1102A, 2021ApJ...920L..43A, 2022SciBu..67.2162D, 2024PhRvD.109l1101A}
and enhancing sensitivity of new physics searches \citep{2022PhRvD.106f3026A,2022SciBu..67..679D}.
For more details on the detector design and on-orbit performance, see \citep{TheDAMPE2017} and \citep{Ambrosi2019}.

The BGO sub-detector is an imaging calorimeter designed to measure particle energies ranging from $\sim$ GeV to $>10$ TeV. The sensitive material consists of 308 BGO crystals, each with dimensions of $60\times2.5\times2.5$
cm$^3$, arranged in 14 orthogonal layers. The BGO has a total thickness of $32X_0$,
which enables nearly full absorption of electromagnetic showers below tens of TeV.
To achieve extended dynamic range, scintillation light from both ends separately undergoes attenuation through two optical filters with 5:1 attenuation ratio, followed by signal readout via three dynodes (dy2, dy5, dy8) of two independent photomultipliers.
With this design, one BGO crystal can measure energy deposition in six overlapping energy
ranges: 2 MeV - 500 MeV, 80 MeV - 20 GeV, 3.2 GeV - 800 GeV from the high-gain end, and
10 MeV - 2.5 GeV, 400 MeV - 100 GeV, 16 GeV - 4 TeV from the low-gain end. The real-time calibration of inter-dynode gain ratios (dy2/dy5 and dy5/dy8) is continuously maintained during operation. Additionally, each front-end electronics (FEE) channel undergoes monthly linearity verification through onboard charge injection circuits (e.g., the DAC calibration) \citep{7312511}.
Furthermore, the BGO fluorescence quenching effect under high energy deposition conditions has been
investigated using a high intensity laser instrument, and no significant nonlinearity
was found \citep{ZHAO2022166453}. Ultimately, a wide dynamic range of $2\times10^6$, a good
and stable linearity up to TeV with a single BGO crystal is achieved across the range from 2 MeV
to 4 TeV. 

The energy scale of each individual BGO bar is calibrated using the ratio of the ADC value
of proton minimum ionization particles (MIP) to the predicted energy by Monte Carlo
(MC) simulations. Geant4 simulations indicate that protons undergoing minimum ionization deposit around 23 MeV\footnote{Note that the mean energy loss rate of
protons ($dE/dx$) depends on particle velocity \citep{pdg2020}. The DAMPE BGO calorimeter, with a thickness of ($32X_{0}$), exhibits a total ionization energy loss exceeding 300MeV in all layers. This results in protons having, on average, $\sim300$ MeV lower kinetic energy in the last layer compared to the first layer. Such significant energy loss leads to $\sim2\%$ variation in $dE/dx$. Consequently, during energy reconstruction, crystals in different layer positions are assigned distinct MIP energies based on simulations.} per BGO bar.
In the FEEs, the energy is read out from dynode 8 at both ends of each crystal. With 308 crystals in total, this configuration yields 616 dy8 ADC-to-Energy ratios. For the energy measurement of one event, the ADC
count from dy2\&dy5 is first converted to dy8 ADC values using dynode ratios, then transformed into energy (in unit of MeV) via the MIP's dy8 ADC-to-Energy ratio \citep{BGOCali}.
The total energy is obtained by summing energies from all bars, incorporating corrections for energy leakage in dead materials \citep{YUE201711}.
To mitigate temporal variations in MIP signals caused by temperature fluctuations, detector aging, solar activities and other factors across different timescales, these ratios are calibrated every orbit when the satellite traverses the latitude region between $-20^\circ$ and $20^\circ$.

With the described calibration procedures, the calorimeter's energy scale is established  based on the detector's MIP response derived from Geant4 simulations. Although this methodology is widely adopted for space-borne particle detector calibration, there are possible
uncertainties from material budget, input proton spectrum, BGO
light yield, readout electronics and so on, which may lead to biases of the absolute
energy scale determination. Therefore, it is important to check the absolute energy
scale of the detector using independent methods. Effective calibration techniques must address two critical requirements. One is to identify a sufficiently sharp spectral structure with well-defined energies. The other is that such a structure should be precisely
measured by the detector. For DAMPE in Low Earth Orbit, the geomagnetic rigidity cutoff\citep[GRC;][]{cutoff1991} on the spectrum of cosmic ray electrons and positrons (CREs) provides an optimal calibration structure for the purpose of the absolute energy scale measurement. The GRC
appears around 10 GeV, where DAMPE achieves $1.5\%$\citep{TheDAMPE2017} energy resolution for electrons/positrons, enabling precise cutoff spectrum measurement. This approach follows precedent established by Fermi-LAT \citep{Ackermann:2011uaq,LAT2TeV} and CALET \citep{PhysRevLett.119.181101} experiments for calorimeter energy scale verification.

In this work, we determine the absolute energy scale of DAMPE's BGO calorimeter by comparing the measured GRC in spectra of CREs with the expected spectra by an algorithm tracing CREs in the geomagnetic field (hereafter backtracing code). The analysis is conducted across 4 distinct McIlwain $L$ intervals: [0,1], [1,1.14], [1.14,1.37], and [1.37,1.64], corresponding to characteristic energies from 7 GeV to 16 GeV. The McIlwain $L$ Parameter is defined as the radial distance (in units of Earth radius) at which a magnetic field line intersects the magnetic equator within a dipolar magnetic field. As described in \cite{SMART20052012}, vertical geomagnetic rigidity cutoffs (Rc) can be approximated by the relation $R_c = 14.6/L^2$ (in units of GV). Due to this direct relationship with cutoff rigidity, the $L$ parameter is highly suitable for cataloging GRC.

\section{Low-energy CREs spectral measurements with DAMPE}\label{sec:Reco}

\subsection{Preselections of the data}

The data used in this work are the events recorded by DAMPE from 2016-04-01 to 2024-07-01.
In total there are about $1.5\times10^{10}$ triggered events. We exclude events
recorded when the satellite travels through the South Atlantic Anomaly (SAA) region, and events collected during geomagnetic storms (Kp$\ge5$) \citep{kpindex}. The latter exclusion aims to reduce the fluctuation
of GRCs which are closely related to the magnitude of the dipole moment of geomagnetic field\citep{SMART20052012}. Events passing the
high-energy trigger and with deposited energies between 2 GeV and 100 GeV are selected.
Side events and bottom-up events are excluded via requiring a reconstructed track across PSD and
the top-four layers of BGO. Heavy ions are also excluded through requiring the PSD charge
measurement to be smaller than 1.55. To select well-contained CREs, events with a maximum
energy deposition in the outermost bar in each of the top-four BGO layers are excluded. After these selections, about $3.73\times10^8$ events remain in the sample.

\subsection{CREs identification}

In the reference\citep{Chang2017b}, CREs were identified with a morphology parameter, $\zeta$,
which combines the lateral extension and longitudinal development of the shower. The parameter
construction and its value for selection are optimized for high-energy CREs identification
($\sim$TeV). However, CREs with energies of tens of GeV deposit almost all of the energy into
the first ten layers of BGO, and the particle identification method should be adjusted
accordingly for such low-energy events. In this work, we develop an alternative particle
identification (PID) parameter based on the three-dimensional shower shape:
\begin{equation}\label{eq:pid}
\begin{split}
{\rm PID} = F(E)\left[\log(R_{r})\sin\theta+\log(R_{l})\cos\theta\right],\\
R_{r}  = \sqrt{{\sum\limits_{i=0}^{n} E_{i}(r_{i}-r_{c})^2}/{\sum\limits_{i=0}^n E_{i}}}\\
R_{l}  = \sqrt{{\sum\limits_{i=0}^{n} E_{i}(l_{i}-l_{c})^2}/{\sum\limits_{i=0}^n E_{i}}}
\end{split}
\end{equation}
where $R_{r}$ and $R_{l}$ are crystal energy weighted radial and longitudinal shower extensions. $ E_{i}$ represents the energy of the i-th fired crystal, $r_{i}-r_{c}$ denotes the shortest distance from the axis of the i-th fired crystal bar to the shower axis, and $l_{i}-l_{c}$ is the distance from the projection point of the i-th fired crystal bar axis onto the shower axis to the center of the shower axis.
The angle $\theta$ is a rotation in the
$R_{r}-R_{l}$ plane that ensures the distribution of $\eta = \log(R_{r})\sin\theta+\log(R_{l})\cos\theta$ for both electrons and hadrons approximates a Gaussian function. The empirical polynomial $F(E)$, obtained by fitting the $\eta$-energy relationship, effectively removes the PID parameter's energy dependence. The misalignment of BGO crystals caused slight shifts in PID peaks. Based on PID peak differences between MC and data, crystal coordinates in MC are corrected. This methodology remains specific to this investigation. Comprehensive crystal alignment requires MIP tracks, though such dedicated calibration lies beyond this paper's scope.

\begin{figure}[!htb]
\centering
\includegraphics[width=0.45\textwidth]{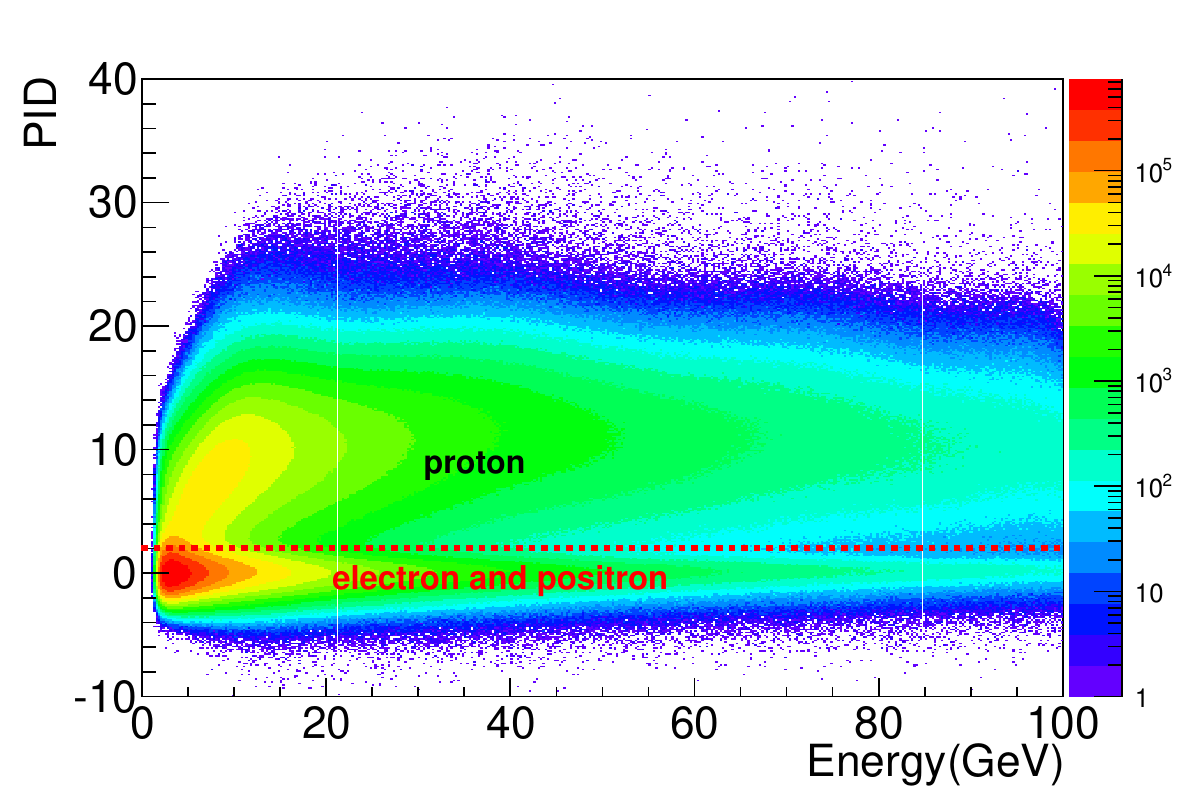}
\caption{The scatter plot of the PID parameter and reconstructed energy of the flight data.
The horizontal dotted line represents ${\rm PID}=2$, which is used to select CREs and reject
hadrons.}
\label{fig:PID}
\end{figure}

Figure ~\ref{fig:PID} shows the scatter plot of PID values versus energy for all preselected events. The horizontal dotted line at ${\rm PID}=2$ distinguishes electron-like events (below the line) from hadronic cosmic rays (mostly protons; above the line).

\subsection{Background estimate}

Two types of background contamination exist in the selected sample: residual hadronic background
and the secondary CREs background \citep{leptonbelowcutoff}. The former arises from the
mis-identification of protons and helium nuclei by the PID selection criteria, while the latter originates from atmospheric interactions of galactic cosmic rays.

\begin{figure}[!htb]
\centering
\includegraphics[width=0.45\textwidth]{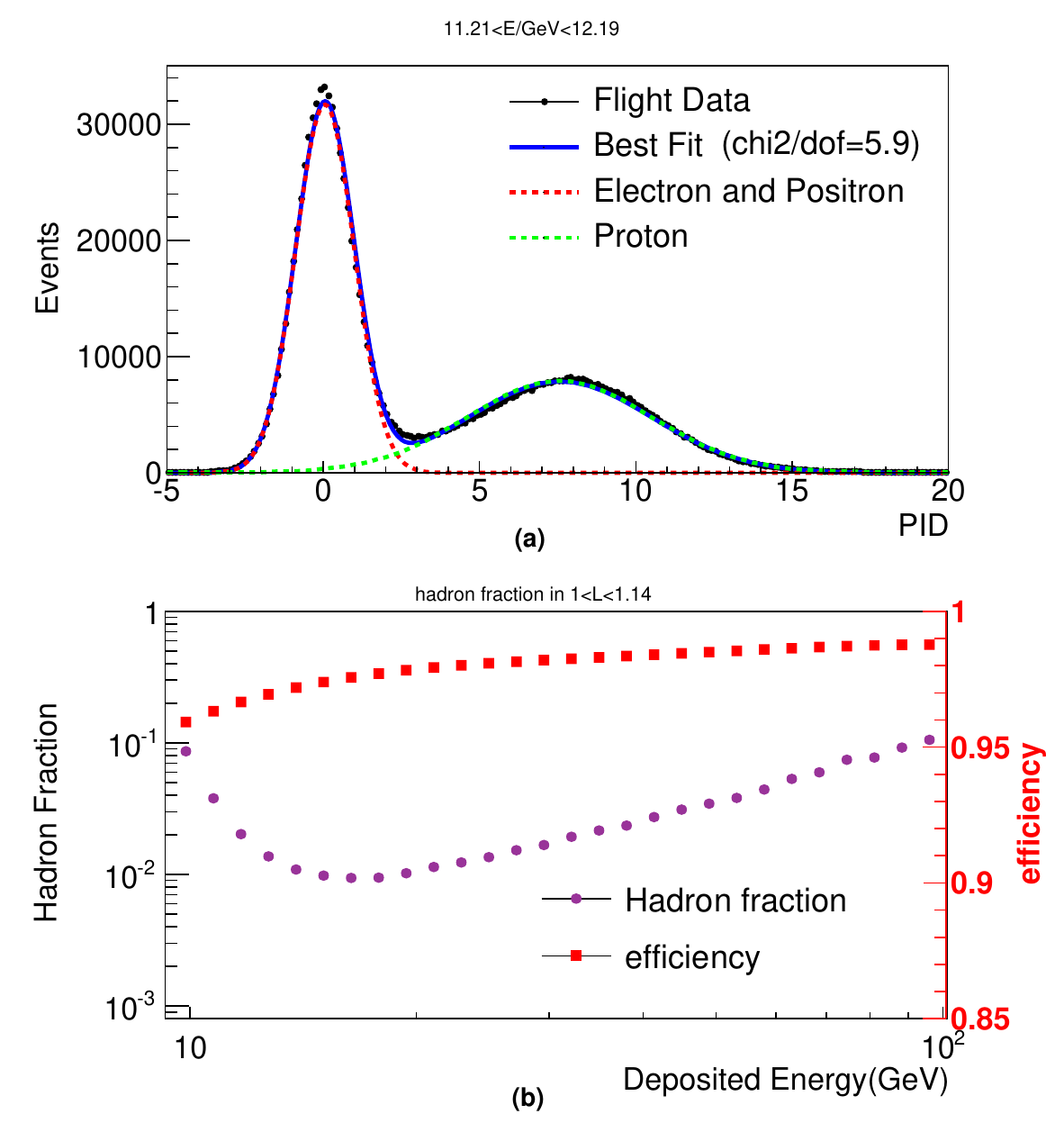}
\caption{(a) Monte Carlo template fitting to the PID distributions in the energy range for $11.2<E/{\rm GeV}<12.2$. Flight data (black dots) are fitted with CREs (red dotted line) and hadronic (green dotted line) templates.
(b) Energy dependence of the hadronic background fraction (purple dots) and electron selection efficiency (red squares) for the $L$ bin $1.00<L<1.14$}.

\label{fig:hadronbackground}
\end{figure}

CREs candidates are obtained by requiring ${\rm PID}<2$. This selection criterion inevitably leads to hadronic contamination from particles with low PID values. To quantify this contamination, we apply a template fitting method to the PID distributions, utilizing MC templates for both CREs and protons. The input spectra for CREs and protons in the MC simulations are weighted according to AMS-02 measurement \citep{AGUILAR20211}.
Although solar modulation effects \citep{solarmodulation} cause slight discrepancies between these spectra and DAMPE observations, the PID parameter's weak energy dependence ensures the templates remain valid.
As demonstrated in the top panel of Figure \ref{fig:hadronbackground}, the MC templates provide a general fit to flight data across all energy and $L$ bins.
The bottom panel of Figure \ref{fig:hadronbackground} displays the energy-dependent hadronic contamination fraction for the
$(1.00<L<1.14)$ bin. The contamination remains below a few percent across most energies while maintaining electron selection efficiency above 95\%. This low contamination level also minimizes secondary proton interference during secondary background estimation. Results for the other three $L$ bins show similar contamination levels.

The interaction of primary particles with the Earth's atmosphere will produce secondary
electrons and positrons, which can enter the field of view (FOV) of DAMPE after being
re-deflected by the geomagnetic field. Unlike hadronic background, these secondary particles cannot be distinguished
from primary CREs using the PID method. However, the azimuth angle distribution of their
arrival direction in the horizontal coordinate system can be used to estimate the fraction of
secondary electrons and positrons, considering the fact that primary
electrons (positrons) come mainly from east (west), while secondaries come more
isotropically \citep{Ackermann:2011uaq}.To model arrival direction, we employ the tracing code described in
\citep{SMART20052012} to trace the trajectories of electrons and positrons in the
geomagnetic field. In this work, we adopt the 13th generation of the International
Geomagnetic Reference Field \citep[IGRF;][]{IGRF13}. Particles are emitted homogeneously
and isotropically on the DAMPE's orbit. Only the events pointing inwards are
traced after reversing the direction of momentum and the polarity of the particle charge. If a particle can reach the place beyond 10 times of the Earth radius where the
geomagnetic field is too weak to bend the particle back, it is considered as a primary
particle. Otherwise, if the particle's trajectory intersects with the Earth surface,
it is discarded. For further details about the tracing strategy, one can refer to \citep{Dai_2022}.

Obviously, backtracing code can only provide distributions (position, direction) of all primary CREs along DAMPE orbit, not the measured distribution. The latter comprises particles that were triggered by the detector, subsequently reconstructed, and passed pre-selection cuts. So,
when the position and angle of primary CREs (passed backtracing code) on DAMPE's orbit is obtained,
it will be used as a particle source connecting to the Geant4 simulation software of DAMPE's detector.

In addition, particles are generated randomly along the DAMPE orbit. This generation process lacks any associated time information. To simulate DAMPE's data acquisition along its actual orbit temporally, we first calculate the event rate of DAMPE using the integral primary CRE flux \citep{AGUILAR20211} multiplied by DAMPE's acceptance. Then, based on this event rate and the operation time of DAMPE, an artificial timestamp is assigned to each simulated event (modelled as a Poisson process with an expectation value of the time interval between events equals to the total number of events divided by the event rate). Crucially, each timestamp corresponds to a specific point on the DAMPE orbit, which defines the particle's position. Thus, the simulated particle positions effectively traverse the DAMPE orbit sequentially point-by-point. From the first particle to the last one, the entire simulated particles will traverse the entire DAMPE orbit point-by-point. Ultimately, the simulation data will be finally produced with a format identical to that of the reconstructed flight data; we term this full on-orbit simulation. 

The primary azimuth template will be constructed
following the same preselection and particle identification procedure as the flight data.
On the other hand, the template of secondaries can be safely extracted from the flight data via requiring
the energy well less than GRC where secondary particles are dominant.

\begin{figure*}[!htb]
\centering
\includegraphics[width=0.95\textwidth]{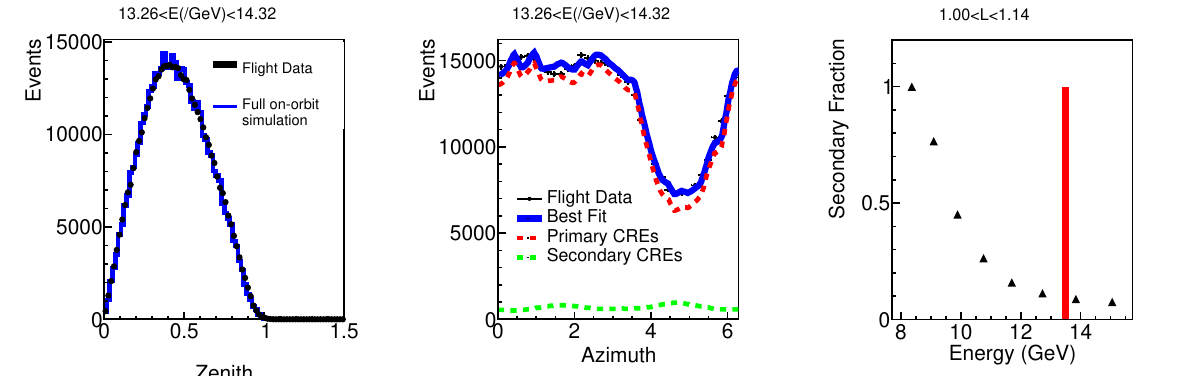}
\caption{Left: the zenith distributions of flight data and full on-orbit simulation data. Zenith distribution of flight data is well described by full on-orbit simulation data. Middle: the azimuth distributions of primaries and secondaries in a typical energy bin near GRC. The East-West effect is well exhibited. Right: secondary fraction versus energy in one $L$ bin of $(1.00,1.14)$. As expected, the secondary fraction near the rigidity cutoff (indicated by the vertical line) is $\sim10\%$. With the decrease of energy, the secondary fraction becomes higher. Situations in the other 3 $L$ bins are similar with the results shown here. The secondary fraction at high-energy part does not drop to zero. The reason is that the hadronic background is not taken into account here.}
\label{fig:azimuth}
\end{figure*}

The Middle panel of Figure \ref{fig:azimuth} displays the azimuth distributions of primaries and
secondaries in a typical energy bin near the rigidity cutoff. The flight data is well fitted
by primary and secondary template. The right panel shows  the energy dependence of the secondary particle fraction for the $L$ bin $(1.00,1.14)$. The secondary fractions are about $\sim10\%$ near the cutoff energies
(marked by a red vertical line). As energies decreases, secondary particles gradually dominant the species. At higher energy side ($E>20$ GeV), the azimuth distributions become nearly isotropic, rendering them ineffective for distinguishing primaries from secondaries. However, primary CREs are expected to dominate in this regime, as the Larmor radius of 20 GeV electron in
Earth's geomagnetic field (typically, $B\sim0.2$ G) exceeds 3300 km,  preventing secondary particles from being magnetically deflected back into the FOV of DAMPE, that is constantly oriented in the direction away from the Earth's center.

Although zenith angle distributions do not have the ability to estimate the fraction of primary components, we still conducted a comparative analysis of the zenith distributions from full on-orbit simulation data and flight data. This approach is motivated by the fact that the GRC measurement inherently integrates over DAMPE’s field of view (FOV), which exhibits a strong zenith-angle dependence due to the spacecraft’s Earth-avoiding pointing strategy. Left panel of Figure \ref{fig:azimuth} compares these distributions in a typical energy bin. The close match between the two datasets confirms identical zenith-angle weighting in both analyses, and further demonstrates that the angular distributions of CREs have been well described by full on-orbit simulation.

\section{Geomagnetic cutoff energy of CREs}

\begin{figure*}[!htb]
\centering
\includegraphics[width=0.85\textwidth]{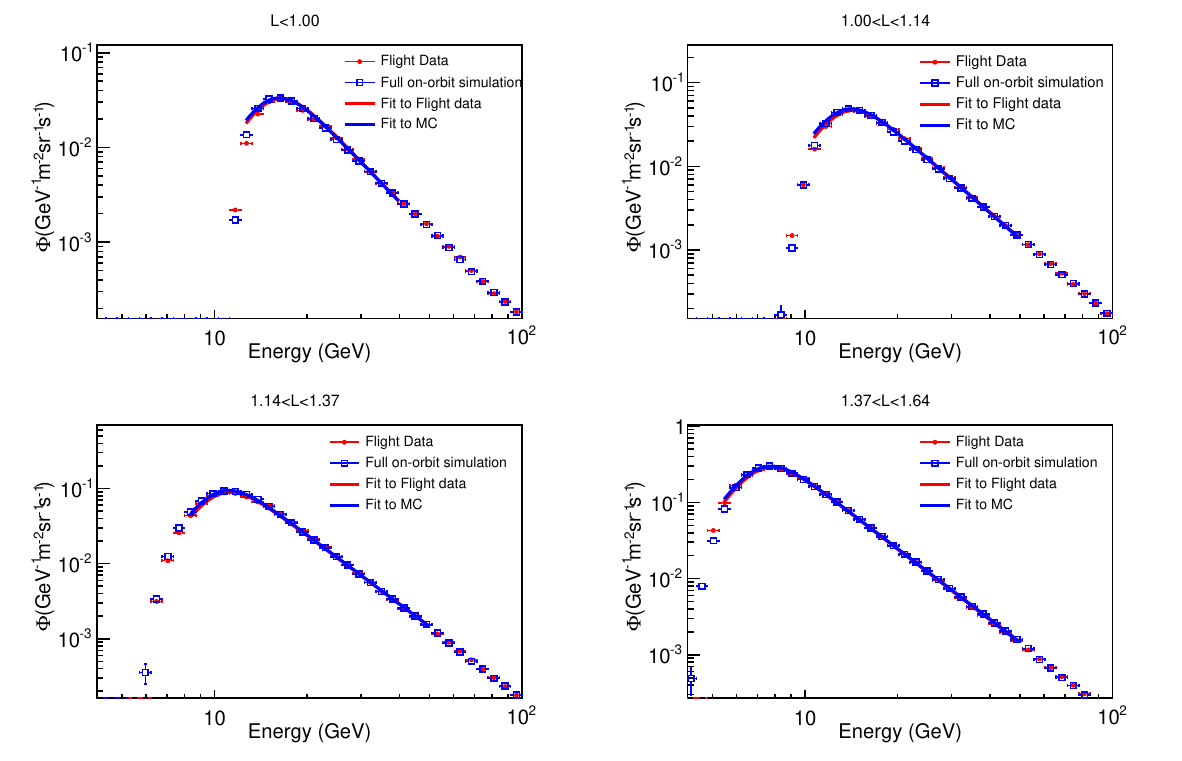}
\caption{Primary CREs fluxes of flight data and full on-orbit simulation in 4 $L$ bins after subtracting hadronic and secondary backgrounds.
Red (blue) dots denote the fluxes from flight data (full on-orbit simulation), the red solid line and the blue dashed line represent the best-fit curves for flight data and full on-orbit simulation, respectively.
}
\label{fig:fluxlbins}
\end{figure*}

The differential flux of primary CREs in each energy bin is calculated by
\begin{equation}\label{eq:flux}
\Phi(E)=\frac{N(1-f_{s})(1-f_{h})}{(A_{\rm eff}T\Delta E)},
\end{equation}
where $N$ is the count of CREs candidates, $f_{s}$ and $f_{h}$ are the fraction of secondary and hadronic background, $A_{\rm eff}$ is the effective acceptance, $\Delta E$ is the energy bin width, and $T$ is the exposure time. The effective acceptance $A_{\rm eff}$ is derived from the MC data, and the exposure time is calculated by summing the duration of all recorded
events and subtracting the dead time of the data acquisition system (DAQ). The differential fluxes in the 4 MacIlwain $L$
bins - $(0.00,1.00)$, $(1.00,1.14)$, $(1.14,1.37)$, and $(1.37,1.64)$ - are shown in Figure \ref{fig:fluxlbins}.

We also derive the expected CREs fluxes from full on-orbit simulation, that is described in Section 2.3. Since the actual CREs flux decreases too quickly with energy
\citep[\bf{$\Phi_{e^-}\sim E^{-3.1}, \Phi_{e^+}\sim E^{-2.9}$};][]{AGUILAR20211}, to accumulate sufficient statistics in all energy bins, simulated CREs are generated with a flat spectrum and then weighted to the DAMPE-measured flux near
the polar region ($L>2.5, GRC<0.6GeV/c$), where primaries dominate above 2GeV. To validate the reliability of the weighting method, we artificially introduced a 2\% energy-scale bias into both the polar-region data and the full on-orbit simulation data. The results demonstrate that this bias was accurately recovered.
The procedure for obtaining full on-orbit simulation fluxes is identical to that for flight data analysis, except no background subtraction is applied
to the simulation data. The full on-orbit simulation fluxes in the same 4 $L$ bins are superimposed in Figure \ref{fig:fluxlbins}.
As expected, the full on-orbit simulation fluxes are well consistent with the flight data above GRC, as the geomagnetic field can hardly alter the flux of high energy cosmic rays. However, below GRC, the differing cutoff energies between flight data and full on-orbit simulation lead to peak position of fluxes at distinct energies, resulting in significant deviations of fluxes between the two datasets.

The absolute energy scale is determined via comparing the spectral cutoffs of flight data
with full on-orbit simulation. We employ a function $\Phi = cE^{\alpha}/(1+E/E_{c})^{-\beta}$ to fit both the
flight data and full on-orbit simulation fluxes, where $\alpha$ is the spectral index, $E_{c}$ is the rigidity cutoff, $\beta$ is an empirical parameter quantifying the flux decline gradient below GRC and here it is fixed at 8.7. The best-fit cutoff energies, $E_c$, for the 4 $L$ bins are listed in Table \ref{tab:grc}, and the best-fit curves are plotted in Figure
\ref{fig:fluxlbins}. The results indicate that the flight data cutoff energies are $0.8\%\sim1.7\%$ higher than those predicted by full on-orbit simulation. On average, the absolute energy scale of the flight data is found to be $1.013\pm0.012$ times higher than full on-orbit simulation predictions over the 7–16 GeV range.

\begin{table*}[htbp]
\begin{center}
\caption{Cutoff energies of flight data $E_c^{\rm data}$ and full on-orbit simulation $E_c^{\rm tracer}$  in 4 $L$ intervals are compared. All errors on the cutoff energies and ratios are statistical only. }
\begin{tabular}{c | c c c }
\hline
$L$ intervals & $E_c^{\rm data}$ & $E_c^{\rm tracer}$ & $E_c^{\rm data}/E_c^{\rm tracer}$ \\
              & (GeV)            & (GeV)              & \\
\hline
$0.00<L<1.00$ & $15.66 \pm 0.36$ & $15.41 \pm 0.39$ & $1.017 \pm 0.034$ \\
$1.00<L<1.14$ & $13.50 \pm 0.27$ & $13.29 \pm 0.26$  & $1.016 \pm 0.028$ \\
$1.14<L<1.37$ & $10.47 \pm 0.15$ & $10.39 \pm 0.16$ & $1.008 \pm 0.021$ \\
$1.37<L<1.64$ & $7.29 \pm 0.10$ & $7.22 \pm 0.11$ & $1.010 \pm 0.021$ \\
\hline
\end{tabular}
\label{tab:grc}
\end{center}
\end{table*}

\section{Systematic uncertainties of the absolute energy scale determination}

One of the major systematic uncertainties comes from the IGRF model and backtracing code.

Given that the uncertainty of the geomagnetic field is about tens of nT, on the order of
$0.1\%$ of the surface geomagnetic field strength \citep{Lowes2000AnEO,Beggan2022}, we
increase or decrease the geomagnetic field by $0.1\%$ and recalculate the GRCs. This leads
to a systematic uncertainty of $\sim1.9\%$ on the cutoff energy.

The spectral index of CREs varies with energy. A function $\Phi = cE^{\alpha}/(1+E/E_{c})^{-\beta}$ is used in the fitting procedure. Consequently, a limited energy range around the cutoff is selected to perform the fit.
To estimate the impact of the selected energy range, we vary the fitting range by adjusting its lower bound to $0.8 \times {\rm GRC}$ and upper bound to $5\times {\rm GRC}$. This results in a relative variation in the cutoff energy of approximately $1.1\%$.

During spectral fitting of the CRE flux, the fitted $E_{c}$ exhibits dependence on the spectral shape parameter $\beta$. To mitigate potential bias from $\beta$ optimization, $\beta$ is fixed at an empirical value of 8.7. This $\beta$ selection may introduce systematic effects in the $E_{c}$ ratio. To quantify $\beta$-induced systematic uncertainty, spectral refitting was performed with $\beta$ as a free parameter. The resulting variation in the $E_{c}$ ratio remains constrained within 1.2\%.

Due to the energy resolution, CREs candidates from one bin may migrate to neighbouring bins, affecting the flux and cutoff energy calculations. To estimate the uncertainty from
the energy resolution, we adopt an unfolding method using the correlation matrix between
incident energy and measured energy from MC simulations. The difference of the cutoff values
before and after unfolding is found to be $\lesssim0.5\%$.

The backgrounds from both hadrons and secondary electrons/positrons are estimated using the template fitting method. For the hadronic background, we randomly vary the fraction values according to a Gaussian function, with $\mu$ and $\sigma$ set to the best-fit fraction values and their uncertainties. We then recalculate the CREs fluxes and repeat the template fitting. A total of 5000 random samplings are performed, yielding a Gaussian-like distribution of GRCs. The systematic uncertainty from the hadronic background is quantified as the standard deviation $\sigma$ relative to the mean $\mu$ of the GRC distribution. In all 4 $L$ bins, uncertainties are found to be less than 0.2\%.

The secondary template is constructed using flight data from a fixed energy range far less than the cutoff.  To assess systematic uncertainties, we vary the energy range by either expanding or halving its width, or shifting the interval boundaries toward lower or higher energies. These modified templates are then applied in the secondary background estimation procedure. The resulting variation in the cutoff energy determination is found to be $\sim0.15\%$.

In summary, the dominant source of systematic uncertainties is from the IGRF model and backtracing code,
which accounts for $1.9\%$. Adding all the above systematic uncertainties
in quadrature, the total systematic error is estimated to be 2.6\%.

\section{Conclusion and discussion}

\begin{figure}[!htb]
\centering
\includegraphics[width=0.45\textwidth]{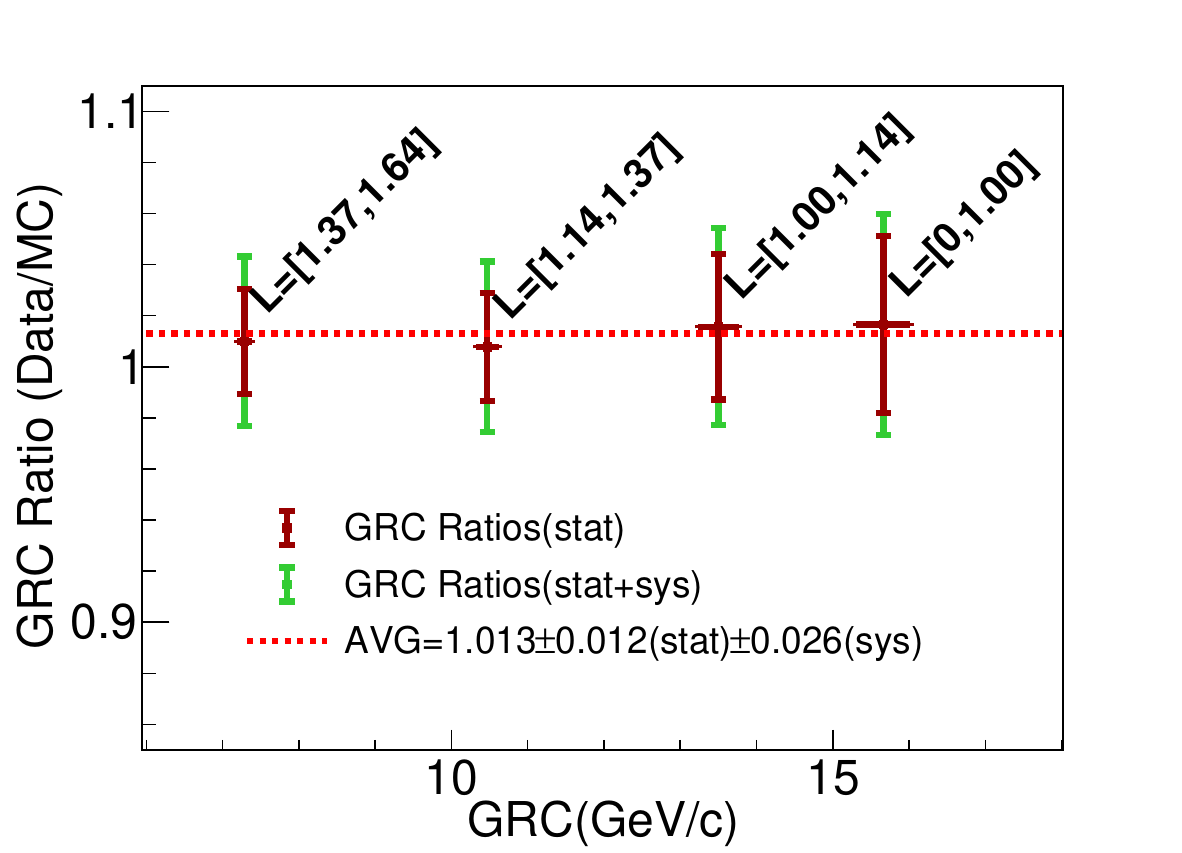}
\caption{Ratios of GRCs of the flight data to the full on-orbit simulation expectation. The errors of GRC are given by spectral fitting, and the red error bars of ratios show the statistical uncertainties only, while the green error bars represent the quadratic sum of the systematic errors and statistical errors. The horizontal dashed line indicates the averaged ratio of $1.013\pm0.012(stat)\pm0.026(sys)$.}
\label{fig:ratio}
\end{figure}

In this work, primary CREs fluxes from 2 GeV to 100 GeV in 4 distinct MacIlwain $L$ regions
are measured to determine the absolute energy scale of the DAMPE calorimeter. Through comparing
the measured spectra and those expected from full on-orbit simulation, the
cutoff energy from the flight data is found to be $1.013\pm0.012(stat)\pm0.026(sys)$ times
higher than full on-orbit simulation result. The absolute energy scale determined for different $L$ bins
varies from $1.008$ to $1.017$, remaining consistent within statistical uncertainties. The relationship between GRC ratios and GRC values is shown in Figure \ref{fig:ratio}. Even though the bias of energy scale is not significant considering the systematic uncertainty, implementing corrections remain essential to ensure energy measurement accuracy. For power-law spectrum with spectral index $\Gamma\approx3$(for CREs), a relative bias b in the absolute energy scale is translated into a rigid shift of spectrum itself by the amount of $(\Gamma-1)b=2.6\%$\citep{Ackermann_2012}. This correction brings DAMPE's CREs spectrum reported in \cite{Chang2017b} into closer agreement with AMS-02 results\citep{AMS02CREs}.

As an independent validation, a measurement of the deposited cosmic iron (Fe) spectrum within 1<L<1.14 GV is reported in Ref\citep{Dai_2022}. Through a similar comparative analysis of Fe spectral cutoffs derived from flight data and ull on-orbit simulation, the study reveals a 1.01 higher absolute energy scale at energies approaching 120 GeV. However, due to the uncertainties of the hadron model, the error is relatively large.

The main reason for the absolute energy scale deviation is likely to be related to the cosmic ray proton MIP energy scale. DAMPE uses the energy deposition of cosmic ray proton MIPs for absolute energy measurement. To minimize the short-term fluctuations in the cosmic ray proton energy spectrum due to geomagnetic disturbances, the on-orbit MIPs trigger mode is enabled only between $20^{\circ}N$ and $20^{\circ}S$. Here, a higher cutoff rigidity helps maintain the primary proton spectrum's stability. However acquired proton MIPs events have a significant fraction of secondary protons. The simulation software uses the AMS02-measured primary proton spectrum\citep{PhysRevLett.114.171103}, combined with the older AMS01-measured secondary proton spectrum\citep{AGUILAR2002331} (as no more recent one is available to the public) as the input particle source. The 1.3\% difference is likely attributable to the uncertainty in the secondary proton spectrum. Currently, the DAMPE collaboration's calibration task force is investigating methods to obtain a more precise secondary proton spectrum through combined software simulations and experimental measurements. The preliminary energy scale of proton MIPs is in general agreement with the result presented in this paper. It is anticipated that the DAMPE collaboration will issue the new version of energy scale corrected data in a near future.

\section*{Acknowledgments}
The DAMPE mission is funded by the strategic priority science and technology projects in space
science of Chinese Academy of Sciences. This work is supported by the National Key Research and Development Program of China (No. 2022YFF0503301), the National Natural Science Foundation of China (No. 12220101003, 12273120), the Strategic Priority Program on Space Science of Chinese Academy of Sciences (No. E02212A02S), the Project for Young Scientists in Basic Research of the Chinese Academy of Sciences (No. YSBR-061), the Youth Innovation Promotion Association of CAS and the Young Elite Scientists Sponsorship Program by CAST (No. YESS20220197).

\bibliographystyle{elsarticle-num}

\bibliography{DAMPE_Absolute_Energy_Scale}

\begin{thebibliography}{10}
\expandafter\ifx\csname url\endcsname\relax
  \def\url#1{\texttt{#1}}\fi
\expandafter\ifx\csname urlprefix\endcsname\relax\def\urlprefix{URL }\fi
\expandafter\ifx\csname href\endcsname\relax
  \def\href#1#2{#2} \def\path#1{#1}\fi

\bibitem{Chang2017b}
{DAMPE Collaboration}, {Direct detection of a break in the teraelectronvolt
  cosmic-ray spectrum of electrons and positrons}, Nature 552~(7683) (2017)
  63--66.
\newblock \href {http://arxiv.org/abs/1711.10981} {\path{arXiv:1711.10981}},
  \href {https://doi.org/10.1038/nature24475} {\path{doi:10.1038/nature24475}}.

\bibitem{2019SciA....5.3793A}
Q.~{An}, et~al, {Measurement of the cosmic ray proton spectrum from 40 GeV to
  100 TeV with the DAMPE satellite}, Science Advances 5~(9) (2019) eaax3793.
\newblock \href {http://arxiv.org/abs/1909.12860} {\path{arXiv:1909.12860}},
  \href {https://doi.org/10.1126/sciadv.aax3793}
  {\path{doi:10.1126/sciadv.aax3793}}.

\bibitem{2021PhRvL.126t1102A}
F.~{Alemanno}, et~al, {Measurement of the Cosmic Ray Helium Energy Spectrum
  from 70 GeV to 80 TeV with the DAMPE Space Mission}, Physical Review Letters
  126~(20) (2021) 201102.
\newblock \href {http://arxiv.org/abs/2105.09073} {\path{arXiv:2105.09073}},
  \href {https://doi.org/10.1103/PhysRevLett.126.201102}
  {\path{doi:10.1103/PhysRevLett.126.201102}}.

\bibitem{2021ApJ...920L..43A}
F.~{Alemanno}, et~al, {Observations of Forbush Decreases of Cosmic-Ray
  Electrons and Positrons with the Dark Matter Particle Explorer},
  Astrophysical Journal Letters 920~(2) (2021) L43.
\newblock \href {http://arxiv.org/abs/2110.00123} {\path{arXiv:2110.00123}},
  \href {https://doi.org/10.3847/2041-8213/ac2de6}
  {\path{doi:10.3847/2041-8213/ac2de6}}.

\bibitem{2022SciBu..67.2162D}
F.~{Alemanno}, et~al, {Detection of spectral hardenings in cosmic-ray
  boron-to-carbon and boron-to-oxygen flux ratios with DAMPE}, Science Bulletin
  67~(21) (2022) 2162--2166.
\newblock \href {http://arxiv.org/abs/2210.08833} {\path{arXiv:2210.08833}},
  \href {https://doi.org/10.1016/j.scib.2022.10.002}
  {\path{doi:10.1016/j.scib.2022.10.002}}.

\bibitem{2024PhRvD.109l1101A}
F.~{Alemanno}, et~al, {Measurement of the cosmic p +He energy spectrum from 50
  GeV to 0.5 PeV with the DAMPE space mission}, Physical Review D 109~(12)
  (2024) L121101.
\newblock \href {http://arxiv.org/abs/2304.00137} {\path{arXiv:2304.00137}},
  \href {https://doi.org/10.1103/PhysRevD.109.L121101}
  {\path{doi:10.1103/PhysRevD.109.L121101}}.

\bibitem{2022PhRvD.106f3026A}
F.~{Alemanno}, et~al, {Search for relativistic fractionally charged particles
  in space}, Physical Review D 106~(6) (2022) 063026.
\newblock \href {http://arxiv.org/abs/2209.04260} {\path{arXiv:2209.04260}},
  \href {https://doi.org/10.1103/PhysRevD.106.063026}
  {\path{doi:10.1103/PhysRevD.106.063026}}.

\bibitem{2022SciBu..67..679D}
{DAMPE Collaboration}, {Search for gamma-ray spectral lines with the DArk
  Matter Particle Explorer}, Science Bulletin 67~(7) (2022) 679--684.
\newblock \href {http://arxiv.org/abs/2112.08860} {\path{arXiv:2112.08860}},
  \href {https://doi.org/10.1016/j.scib.2021.12.015}
  {\path{doi:10.1016/j.scib.2021.12.015}}.

\bibitem{TheDAMPE2017}
J.~{Chang}, et~al, {The DArk Matter Particle Explorer mission}, Astroparticle
  Physics 95 (2017) 6--24.
\newblock \href {http://arxiv.org/abs/1706.08453} {\path{arXiv:1706.08453}},
  \href {https://doi.org/10.1016/j.astropartphys.2017.08.005}
  {\path{doi:10.1016/j.astropartphys.2017.08.005}}.

\bibitem{Ambrosi2019}
G.~{Ambrosi}, et~al, {The on-orbit calibration of DArk Matter Particle
  Explorer}, Astroparticle Physics 106 (2019) 18--34.

\bibitem{7312511}
C.~Feng, et~al, {Design of the Readout Electronics for the BGO Calorimeter of
  DAMPE Mission}, IEEE Transactions on Nuclear Science 62~(6) (2015)
  3117--3125.
\newblock \href {https://doi.org/10.1109/TNS.2015.2479091}
  {\path{doi:10.1109/TNS.2015.2479091}}.

\bibitem{ZHAO2022166453}
C.~Zhao, , et~al, {The study of fluorescence response to energy deposition in
  the BGO calorimeter of DAMPE}, Nuclear Instruments and Methods in Physics
  Research Section A: Accelerators, Spectrometers, Detectors and Associated
  Equipment 1029 (2022) 166453.
\newblock \href {https://doi.org/https://doi.org/10.1016/j.nima.2022.166453}
  {\path{doi:https://doi.org/10.1016/j.nima.2022.166453}}.

\bibitem{pdg2020}
Particle\_Data\_Group, \href{https://doi.org/10.1093/ptep/ptaa104}{{Review of
  Particle Physics}}, Progress of Theoretical and Experimental Physics 2020~(8)
  (2020) 083C01.
\newblock \href {https://doi.org/10.1093/ptep/ptaa104}
  {\path{doi:10.1093/ptep/ptaa104}}.
\newline\urlprefix\url{https://doi.org/10.1093/ptep/ptaa104}

\bibitem{BGOCali}
Z.~Zhang, et~al, {The calibration and electron energy reconstruction of the BGO
  ECAL of the DAMPE detector}, Nuclear Instruments and Methods in Physics
  Research Section A: Accelerators, Spectrometers, Detectors and Associated
  Equipment 836 (08 2016).
\newblock \href {https://doi.org/10.1016/j.nima.2016.08.015}
  {\path{doi:10.1016/j.nima.2016.08.015}}.

\bibitem{YUE201711}
C.~Yue, , et~al, {A parameterized energy correction method for electromagnetic
  showers in BGO-ECAL of DAMPE}, Nuclear Instruments and Methods in Physics
  Research Section A: Accelerators, Spectrometers, Detectors and Associated
  Equipment 856 (2017) 11--16.
\newblock \href {https://doi.org/https://doi.org/10.1016/j.nima.2017.03.013}
  {\path{doi:https://doi.org/10.1016/j.nima.2017.03.013}}.

\bibitem{cutoff1991}
D.~J. {Cooke}, et~al, {On cosmic-ray cut-off terminology.}, Nuovo Cimento C
  Geophysics Space Physics C 14 (1991) 213--234.
\newblock \href {https://doi.org/10.1007/BF02509357}
  {\path{doi:10.1007/BF02509357}}.

\bibitem{Ackermann:2011uaq}
M.~Ackermann, et~al., {In-flight measurement of the absolute energy scale of
  the Fermi Large Area Telescope}, Astropart. Phys. 35 (2012) 346--353.
\newblock \href {http://arxiv.org/abs/1108.0201} {\path{arXiv:1108.0201}},
  \href {https://doi.org/10.1016/j.astropartphys.2011.10.007}
  {\path{doi:10.1016/j.astropartphys.2011.10.007}}.

\bibitem{LAT2TeV}
S.~Abdollahi, et~al, {Cosmic-ray electron-positron spectrum from 7 GeV to 2 TeV
  with the Fermi Large Area Telescope}, Physical Review D 95 (04 2017).
\newblock \href {https://doi.org/10.1103/PhysRevD.95.082007}
  {\path{doi:10.1103/PhysRevD.95.082007}}.

\bibitem{PhysRevLett.119.181101}
O.~Adriani, et~al,
  \href{https://link.aps.org/doi/10.1103/PhysRevLett.119.181101}{{Energy
  Spectrum of Cosmic-Ray Electron and Positron from 10 GeV to 3 TeV Observed
  with the Calorimetric Electron Telescope on the International Space
  Station}}, Phys. Rev. Lett. 119 (2017) 181101.
\newblock \href {https://doi.org/10.1103/PhysRevLett.119.181101}
  {\path{doi:10.1103/PhysRevLett.119.181101}}.
\newline\urlprefix\url{https://link.aps.org/doi/10.1103/PhysRevLett.119.181101}

\bibitem{SMART20052012}
D.~Smart, M.~Shea, {A review of geomagnetic cutoff rigidities for
  earth-orbiting spacecraft}, Advances in Space Research 36~(10) (2005)
  2012--2020, solar Wind-Magnetosphere-Ionosphere Dynamics and Radiation
  Models.
\newblock \href {https://doi.org/https://doi.org/10.1016/j.asr.2004.09.015}
  {\path{doi:https://doi.org/10.1016/j.asr.2004.09.015}}.

\bibitem{kpindex}
J.~Matzka, et~al,
  \href{https://agupubs.onlinelibrary.wiley.com/doi/abs/10.1029/2020SW002641}{{The
  Geomagnetic Kp Index and Derived Indices of Geomagnetic Activity}}, Space
  Weather 19~(5) (2021) e2020SW002641, e2020SW002641 2020SW002641.
\newblock \href
  {http://arxiv.org/abs/https://agupubs.onlinelibrary.wiley.com/doi/pdf/10.1029/2020SW002641}
  {\path{arXiv:https://agupubs.onlinelibrary.wiley.com/doi/pdf/10.1029/2020SW002641}},
  \href {https://doi.org/https://doi.org/10.1029/2020SW002641}
  {\path{doi:https://doi.org/10.1029/2020SW002641}}.
\newline\urlprefix\url{https://agupubs.onlinelibrary.wiley.com/doi/abs/10.1029/2020SW002641}

\bibitem{leptonbelowcutoff}
J.~Alcaraz, et~al, {Leptons in near earth orbit}, Physics Letters B 484~(1)
  (2000) 10--22.
\newblock \href {https://doi.org/https://doi.org/10.1016/S0370-2693(00)00588-8}
  {\path{doi:https://doi.org/10.1016/S0370-2693(00)00588-8}}.

\bibitem{AGUILAR20211}
M.~Aguilar, et~al, {The Alpha Magnetic Spectrometer (AMS) on the international
  space station: Part II - Results from the first seven years}, Physics Reports
  894 (2021) 1--116, the Alpha Magnetic Spectrometer (AMS) on the International
  Space Station: Part II - Results from the First Seven Years.
\newblock \href {https://doi.org/https://doi.org/10.1016/j.physrep.2020.09.003}
  {\path{doi:https://doi.org/10.1016/j.physrep.2020.09.003}}.

\bibitem{solarmodulation}
M.~S. {Potgieter}, {Solar Modulation of Cosmic Rays}, Living Reviews in Solar
  Physics 10~(1) (2013) 3.
\newblock \href {http://arxiv.org/abs/1306.4421} {\path{arXiv:1306.4421}},
  \href {https://doi.org/10.12942/lrsp-2013-3}
  {\path{doi:10.12942/lrsp-2013-3}}.

\bibitem{IGRF13}
P.~Alken, , et~al, {International Geomagnetic Reference Field: the thirteenth
  generation}, EARTH PLANETS AND SPACE 73~(1) (FEB 11 2021).
\newblock \href {https://doi.org/10.1186/s40623-020-01288-x}
  {\path{doi:10.1186/s40623-020-01288-x}}.

\bibitem{Dai_2022}
H.-T. Dai, et~al, \href{https://dx.doi.org/10.1088/1674-4527/ac47a8}{{Method of
  Separating Cosmic-Ray Positrons from Electrons in the DAMPE Experiment}},
  Research in Astronomy and Astrophysics 22~(3) (2022) 035012.
\newblock \href {https://doi.org/10.1088/1674-4527/ac47a8}
  {\path{doi:10.1088/1674-4527/ac47a8}}.
\newline\urlprefix\url{https://dx.doi.org/10.1088/1674-4527/ac47a8}

\bibitem{Lowes2000AnEO}
F.~J. Lowes, \href{https://api.semanticscholar.org/CorpusID:54542247}{{An
  estimate of the errors of the IGRF/DGRF fields 1945–2000}}, Earth, Planets
  and Space 52 (2000) 1207--1211.
\newline\urlprefix\url{https://api.semanticscholar.org/CorpusID:54542247}

\bibitem{Beggan2022}
C.~D. Beggan,
  \href{https://earth-planets-space.springeropen.com/articles/10.1186/s40623-022-01572-y#citeas}{{Evidence-based
  uncertainty estimates for the International Geomagnetic Reference Field}},
  Earth, Planets and Space 74 (01 2022).
\newline\urlprefix\url{https://earth-planets-space.springeropen.com/articles/10.1186/s40623-022-01572-y#citeas}

\bibitem{Ackermann_2012}
M.~Ackermann, et~al, \href{https://dx.doi.org/10.1088/0067-0049/203/1/4}{{THE
  FERMI LARGE AREA TELESCOPE ON ORBIT: EVENT CLASSIFICATION, INSTRUMENT
  RESPONSE FUNCTIONS, AND CALIBRATION}}, The Astrophysical Journal Supplement
  Series 203~(1) (2012) 4.
\newblock \href {https://doi.org/10.1088/0067-0049/203/1/4}
  {\path{doi:10.1088/0067-0049/203/1/4}}.
\newline\urlprefix\url{https://dx.doi.org/10.1088/0067-0049/203/1/4}

\bibitem{AMS02CREs}
M.~Aguilar, et~al, {Precision Measurement of the
  $({e}^{+}+{e}^{\ensuremath{-}})$ Flux in Primary Cosmic Rays from 0.5 GeV to
  1 TeV with the Alpha Magnetic Spectrometer on the International Space
  Station}, Phys. Rev. Lett. 113 (2014) 221102.
\newblock \href {https://doi.org/10.1103/PhysRevLett.113.221102}
  {\path{doi:10.1103/PhysRevLett.113.221102}}.

\bibitem{PhysRevLett.114.171103}
M.~Aguilar, et~al,
  \href{https://link.aps.org/doi/10.1103/PhysRevLett.114.171103}{{Precision
  Measurement of the Proton Flux in Primary Cosmic Rays from Rigidity 1 GV to
  1.8 TV with the Alpha Magnetic Spectrometer on the International Space
  Station}}, Phys. Rev. Lett. 114 (2015) 171103.
\newblock \href {https://doi.org/10.1103/PhysRevLett.114.171103}
  {\path{doi:10.1103/PhysRevLett.114.171103}}.
\newline\urlprefix\url{https://link.aps.org/doi/10.1103/PhysRevLett.114.171103}

\bibitem{AGUILAR2002331}
M.~Aguilar, et~al, {The Alpha Magnetic Spectrometer (AMS) on the International
  Space Station: Part II results from the test flight on the space shuttle},
  Physics Reports 366~(6) (2002) 331--405.
\newblock \href {https://doi.org/https://doi.org/10.1016/S0370-1573(02)00013-3}
  {\path{doi:https://doi.org/10.1016/S0370-1573(02)00013-3}}.

\end{thebibliography}

\end{document}